\documentclass[conference,a4paper]{IEEEtran}

\usepackage[utf8]{inputenc} 
\usepackage[T1]{fontenc}
\usepackage{url}              

\usepackage[cmex10]{amsmath}  
\usepackage{mleftright}       
\mleftright                   

\usepackage{booktabs}         

\usepackage{amsmath,amssymb,dsfont,stfloats,color,url,bbm}
\usepackage[pdftex]{graphicx}
\usepackage[caption=false,font=footnotesize]{subfig}
\usepackage{mathabx}
\usepackage{multirow}
\usepackage{tabu}

\usepackage{siunitx}
\usepackage{tikz} 
\usepackage{pgfplots}
\usepackage{enumitem}

\pgfplotsset{compat=newest} 
\usepgfplotslibrary{units} 
\DeclareGraphicsExtensions{.eps,.pdf,.png,.jpg,.gif,.jpeg,.pstex}

\usepackage[noadjust]{cite}
\usepackage{url}

\newcommand*{\Resize}[2]{\resizebox{#1}{!}{$#2$}}%

\newfont{\bbb}{msbm10 scaled 700}

\newfont{\bb}{msbm10 scaled 1100}
\newcommand{\CC}{\mbox{\bb C}}
\newcommand{\PP}{\mbox{\bb P}}

\newcommand{\EE}{\mbox{\bb E}}

\newcommand{\HH}{\mbox{\bb H}}

\newcommand{\yy}{\mathbbm{y}}

\newcommand{\zz}{\mathbbm{z}}
\newcommand{\sss}{\mathbbm{s}}

\newcommand{\hh}{\mathbbm{h}}

\newcommand{\vvv}{\mathbbm{v}}

\usepackage[mathscr]{euscript}
\newcommand{\Rs}{\mathscr{R}}
\newcommand{\Cs}{\mathscr{C}}

\newcommand{\av}{{\bf a}}

\newcommand{\hv}{{\bf h}}

\newcommand{\vv}{{\bf v}}
\newcommand{\xv}{{\bf x}}

\newcommand{\zerov}{{\bf 0}}

\newcommand{\Fm}{{\bf F}}

\newcommand{\Qm}{{\bf Q}}

\newcommand{\Ac}{{\cal A}}

\newcommand{\Cc}{{\cal C}}

\newcommand{\Ec}{{\cal E}}

\newcommand{\Gc}{{\cal G}}

\newcommand{\Ic}{{\cal I}}

\newcommand{\Nc}{{\cal N}}

\newcommand{\Sc}{{\cal S}}

\newcommand{\Uc}{{\cal U}}

\newcommand{\nuv}{\hbox{\boldmath$\nu$}}
\newcommand{\muv}{\hbox{\boldmath$\mu$}}

\renewcommand{\arg}{{\hbox{arg}}}

\newcommand{\eqdef}{\stackrel{\Delta}{=}}
\newcommand{\defines}{{\,\,\stackrel{\scriptscriptstyle \bigtriangleup}{=}\,\,}}

\newcommand{\herm}{{\sf H}}

\newcommand{\SINR}{{\sf SINR}}
\newcommand{\SNR}{{\sf SNR}}

\newcommand{\taudmrs}{\tau_p}

\newcommand{\Ktot}{K} 
\newcommand{\Kact}{K_{\rm act}}

\usepackage{stmaryrd} 

\renewcommand{\arg}{{\rm arg}}

\pgfplotsset{
	kurze Legende/.style={%
		legend image code/.code={
			\draw[##1,line width=0.6pt]
			plot coordinates {
				(0cm,0cm)
				(0.25cm,0cm)
			};%
		}
	}
}

\pgfplotsset{
	sehr kurze Legende/.style={%
		legend image code/.code={
			\draw[##1,line width=0.7pt]
			plot coordinates {
				(0cm,0cm)
				(0.17cm,0cm)
			};%
		}
	}
}
\begin{document}
	
	\title{User-Centric Clustering Under Fairness Scheduling in Cell-Free Massive MIMO} 
	
	\author{\IEEEauthorblockN{Fabian G\"ottsch\IEEEauthorrefmark{1},
		Noboru Osawa\IEEEauthorrefmark{2}, Takeo Ohseki\IEEEauthorrefmark{2}, Yoshiaki Amano\IEEEauthorrefmark{2}, \\ Issei Kanno\IEEEauthorrefmark{2}, Kosuke Yamazaki\IEEEauthorrefmark{2}, Giuseppe Caire\IEEEauthorrefmark{1}}
	\IEEEauthorblockA{\IEEEauthorrefmark{1}Technical University of Berlin, Germany\\
		\IEEEauthorrefmark{2}KDDI Research Inc., Japan\\
		Emails: \{fabian.goettsch, caire\}@tu-berlin.de, \{nb-oosawa, ohseki, yo-amano, is-kanno, ko-yamazaki\}@kddi.jp}}
	
	\maketitle

	\maketitle
	
	\begin{abstract}
		We consider fairness scheduling in a user-centric cell-free massive MIMO network, where $L$ remote radio units, each with $M$ antennas, serve $\Ktot \approx LM$ user equipments (UEs). Recent results show that the maximum network sum throughput is achieved where $\Kact \approx \frac{LM}{2}$ UEs are simultaneously active in any given time-frequency slots. However, the number of users $\Ktot$ in the network is usually much larger. This requires that users are scheduled over the time-frequency resource and achieve a certain throughput rate as an average over the slots. We impose throughput fairness among UEs with a scheduling approach aiming to maximize a concave component-wise non-decreasing network utility function of the per-user throughput rates. In cell-free user-centric networks, the pilot and cluster assignment is usually done for a given set of active users. Combined with fairness scheduling, this requires pilot and cluster reassignment at each scheduling slot, involving an enormous overhead of control signaling exchange between network entities.  We propose a fixed pilot and cluster assignment scheme (independent of the scheduling decisions), which outperforms the baseline method in terms of UE throughput, while requiring much less control information exchange between network entities. 
	\end{abstract}
	
	
	\section{Introduction} 
	Cell-free massive MIMO is a form of distributed massive MIMO that has attracted a great deal of interest in industry and research in recent years in order to serve a large number of user equipments (UEs) in dense beyond 5G networks. It is based on multiuser/massive MIMO \cite{caire2003achievable, 3gpp38211, marzetta2010noncooperative}, where the $LM$ access point antennas are distributed across the network area on $L$ remote radio units (RUs), each equipped with $M$ antennas. A research direction towards a practical cell-free network considers a user-centric scalable system \cite{9336188, 9064545}, where user-centric clusters of RUs serve each UE $k \in [K]$.\footnote{The set of integers from 1 to $N$ is denoted by $[N]$.} 
	Due to the distribution of RUs across the network area, cell-free massive MIMO is expected to serve all UEs with approximately the same quality of service. Unfortunately, this is not easy to achieve, even with efforts to make the network more fair \cite{9973365}. 
	While early works on cell-free massive MIMO assumed $L  > K$ \cite[Ch. 2]{9336188} leading to overall large UE data rates, recent works considered the more realistic UE density regime $K > L$, where $K$ is comparable to $ML$ (i.e., in the same order of magnitude) taking into account multi-antenna RUs. This regime yields a relatively unfair distribution of the UE data rates \cite{9973365, gottsch2022subspace}. 
	\looseness=-1
	For cell-free massive MIMO, very high-density scenarios are envisaged (e.g., see the real-world deployment in \cite{7421132}).
	Since the total number of UEs $K$ may be on the order of tens of thousands,  
	it is clear that spatial multiplexing alone cannot support all UEs at the same time. 
	Hence, for $\Ktot$ significantly larger than $ML$, users must be scheduled in the time/frequency domain on different slots, 
	such that on each ``resource block'' (RB), i.e., the slots defining the time-frequency granularity of the scheduler, 
	only  a number $\Kact$ of ``active'' users is served using spatial multiplexing. 
	In particular, recent results  \cite{gottsch2022fairness} have shown that for typical network layouts and operating conditions the network total spectral efficiency (SE) is maximum when $\Kact \approx \frac{LM}{2}$, and such maximum is quite ``flat'', 
	i.e., quite insensitive with respect to the exact value of $\Kact$. 
	\looseness=-1
	
	To give an idea, consider a network with 100 RBs per time slot serving $\Ktot = 10000$ users with $L = 20$ RUs and $M = 16$. Each user is allocated a block of $F = 10$ RBs in frequency to achieve a certain level of frequency diversity. Hence, a scheduler must choose on every RB a set of $\Kact \approx 160$ users out of 1000 users per RB. 
	Therefore, the relevant performance metric is the per-user {\em throughput rate}, 
	i.e., the rate averaged over a long sequence of slots. Since the number of active users is much less than $\Ktot$, it is important to operate the network such that each user obtains a ``fair share'' of the total throughput rate. Hence, 
	the scheduler must be designed to achieve some desired form of fairness of the per-user throughput distribution. 
	Finally, also as a consequence of this setting, we notice that the familiar {\em ergodic rate} 
	used as performance metric in most standard literature on cell-free networks (e.g., see  \cite{9336188, 9064545,9973365, gottsch2022subspace}) is not relevant any longer. 
	In fact, the scheduler must allocate an ``instantaneous'' rate on each RB (or block of $F$ RBs) and decoding is performed block by block, such that averaging over a virtually infinite sequence of fading states is no longer possible.  
	In this case, the instantaneous rate must be scheduled according to the notion of {\em information outage rate} (e.g., see \cite{shirani2010mimo}).  
	\looseness=-1
	\subsection{Related Literature}
	
	Scheduling in cell-free massive MIMO has been considered in relatively few works \cite{chen2019dynamic, ammar2021distributed, denis2021improving, gottsch2022fairness} in comparison with the very large number of papers 
	considering the ergodic rate of a fixed set of ``always active'' users (see, e.g., \cite{9336188} and references therein). 
	We build on the system proposed in \cite{gottsch2022fairness}, which considers $\Ktot > \Kact$ UEs, and sets $\Kact$ as the number of active users that (approximately) maximizes the network total SE.  
	Using the dynamic scheduling framework of \cite{georgiadis2006resource}, hard and proportional fairness scheduling 
	(resp., HFS and PFS) are addressed.  A system design challenge with a high UE density consists of the assignment of 
	uplink (UL) pilots for channel estimation and the user-centric cluster formation \cite{9336188,gottsch2022subspace}. 
	This problem is exacerbated in the presence of dynamic scheduling, since the set of active user changes at every scheduling decision (time slot).  In \cite{gottsch2022fairness}, this challenge is addressed by performing the pilot and cluster assignment at each scheduling decision, i.e., after selecting the set of $\Kact < \Ktot$ active UEs. In fact, users sharing the same pilot and located in proximity of each other may suffer from severe pilot contamination. Nevertheless, pilot contamination comes only from active users, the identity of which is not known prior the scheduling decision on the current slot. 
	In practice, performing these operations at each slot comes at the cost of a large communication overhead between UEs and RUs.
	
	\subsection{Contributions}
	
	We propose a less communication-intensive pilot and cluster assignment scheme with pilots 
	and clusters permanently assigned to the $\Ktot$ UEs. 
	For those co-pilot users that may cause severe mutual pilot contamination, we construct a conflict graph 
	that prohibits these UEs from being scheduled on the same RB. The proposed method requires far less communication and it is much better suited for a practical implementation, compared to the per-slot reassignment scheme in \cite{gottsch2022fairness}.
	Interestingly, our simulations show that it can also achieve a (slightly) larger throughput rate, due to 
	the avoidance of strong co-pilot interference expressed by the conflict graph.
	
	Furthermore, as anticipated before, we use information outage rates for the instantaneous rate scheduling, reflecting the fact that
	channel coding is performed on a block by block basis on a finite number of RBs. 
	In particular, our numerical results show the benefit of a moderate frequency diversity order of $F > 1$ RBs. As $F$ increases, the instantaneous mutual information distribution ``concentrates'' and behaves in a more deterministic way, 
	allowing a more aggressive instantaneous rate allocation on the active slots.\footnote{We consider scheduling over time and allocate all $F$ RBs to active users. The allocation of RBs to different users in the same time slot to avoid conflicts is beyond the scope of this work. 
	The proposed conflict graph-based scheme would be an approach to assign users to different RBs if it is done in a sequential manner, e.g., RB by RB. The problem would then be very similar to what is done in this paper.}
	
	\section{System Description} 
	
	We consider a cell-free massive MIMO network as in \cite{gottsch2022fairness} in TDD operation mode with $L$ RUs, each equipped with $M$ antennas, 
	and $\Ktot$ single-antenna UEs. Both RUs and UEs are distributed on a squared region on the 2-dimensional plane. 
	We let $\HH (t, f) \in \CC^{LM \times \Ktot}$ denote the channel matrix between all the $\Ktot$ UE antennas and all the $LM$ 
	RU antennas on a given RB $f$ in time slot $t$, formed by $M \times 1$ blocks $\hv_{\ell,k}(t, f)$ in correspondence of the $M$ antennas of RU $\ell$ 
	and UE $k$. 
	Letting $\Ac(t)\subseteq [\Ktot]$ denote the set of active users scheduled in slot $t$, 
	the columns $\hh_k(t, f)$ of $\HH (t, f)$ corresponding to inactive UEs $k \in [\Ktot]: k \notin \Ac(t)$ contain the identically zero vector $\zerov$.
	
	Let $\Fm$ denote the $M \times M$ unitary DFT matrix with $(m,n)$-elements
	$\left[ \Fm \right]_{m,n} = \frac{e^{-j\frac{2\pi}{M} mn}}{\sqrt{M}}$ for  $m, n  = 0,1,\ldots, M-1$, and consider the angular support set $\Sc_{\ell,k} \subseteq \{0,\ldots, M-1\}$ 
	obtained according to the single ring local scattering model \cite{adhikary2013joint}, where $\Sc_{\ell,k}$ contains the DFT quantized angles (multiples of $2\pi/M$) falling inside an interval of length $\Delta$ placed symmetrically around the direction joining UE $k$ and RU $\ell$. Then, the channel between RU $\ell$ and the active UE $k$ with large-scale fading coefficient (LSFC) $\beta_{\ell,k}$ on RB $f$ in slot $t$ is
	$\hv_{\ell,k} (t, f) = \sqrt{\frac{\beta_{\ell,k} M}{|\Sc_{\ell,k}|}}  \Fm_{\ell,k} \nuv_{\ell, k} (t, f)$,
	where, using a MATLAB-like notation, $\Fm_{\ell,k} \eqdef \Fm(: , \Sc_{\ell,k})$ denotes the tall unitary matrix obtained by selecting the columns 
	of $\Fm$ corresponding to the index set $\Sc_{\ell,k}$\footnote{Note that for uniform linear arrays (ULAs) and uniform planar arrays (UPAs), as widely used in today's massive MIMO implementations, the channel covariance matrix is Toeplitz (for ULA) or Block-Toeplitz (for UPA), and that large Toeplitz and block-Toeplitz matrices are approximately diagonalized by DFTs on the columns and on the rows (see \cite{adhikary2013joint} for a precise statement based on Szeg\"o's theorem).}, and $\nuv_{\ell,k} (t, f)$ is an $|\Sc_{\ell,k}| \times 1$ i.i.d. Gaussian vector with components 
	$\sim \Cc\Nc(0,1)$. Note that the LSFC, angular support and thus also $\Fm_{\ell,k}$ are independent of the RB and time indices, while $\nuv_{\ell,k} (t, f)$ has different realizations on each RB $f$ and in each slot $t$. As in most cell-free massive MIMO literature (see \cite{9336188}), we assume that the LSFCs are known at the RUs.

	In slot $t$ and for all RBs $f \in [F]$, each UE $k$ is connected to a cluster $\Cc_k(t) \subseteq [L]$ of RUs 
	and each RU $\ell$ has a set of associated UEs $\Uc_\ell(t) \subseteq [\Ktot]$. The UE-RU association is described by 
	a bipartite graph $\Gc(t)$ with two classes of nodes (UEs and RUs) such that the neighborhood of UE-node $k$ is $\Cc_k(t)$ 
	and the neighborhood of RU-node $\ell$ is $\Uc_\ell(t)$. The set of edges of $\Gc(t)$ is denoted by $\Ec(t)$, i.e., $\Gc(t) = \Gc([L], [\Ktot], \Ec(t))$. 
	We assume OFDM modulation and that the channel in the time-frequency domain follows the
	standard block-fading model \cite{marzetta2010noncooperative,9336188,9064545}. The channel vectors from UEs to RUs are random but constant over coherence blocks of 
	$T$ signal dimensions in the time-frequency domain, of which $\taudmrs$ dimensions are used for the finite-dimensional UL pilot signal, such that  a fraction $1-\frac{\tau_p}{T}$ of dimensions per RB is used for data transmission. We assume that one time-frequency slot, i.e., one realization of $t$ and $f$, corresponds to one channel coherence block.
	\looseness=-1
	
	\subsection{Uplink Decoding with Partial Channel State Information}
	
	We consider \textit{partial} channel state information obtained by subspace projection channel estimation. Each RU $\ell$ computes locally the channel estimates $\widehat{\hv}_{\ell,k}(t, f)$ for UEs $k \in \Uc_\ell(t)$ from the received {\em orthogonal} UL pilot signal, where perfect subspace knowledge is assumed (see \cite{gottsch2022subspace} for details).
	
	Based on the channel estimates $\{\widehat{\hv}_{\ell,k} (t, f):  k \in \Uc_\ell(t)\}$,
	RU $\ell$ locally computes a unique receiver combining vector $\vv_{\ell,k} (t, f)$ for each associated UE $k \in \Uc_\ell$, where a linear MMSE principle is used. For $k \notin \Uc_\ell(t)$, we have $\vv_{\ell,k} (t, f) = \zerov$.
	The cluster $\Cc_k(t)$ combines the vectors $\{\vv_{\ell,k} (t, f) : \ell \in \Cc_k(t)\}$ to form a receiver {\em unit norm} vector $\vvv_k (t, f) \in \CC^{LM \times 1}$ for UE $k$, aiming to maximize the UL signal to Interference plus noise ratio (SINR) (see \cite{9593169} for details). Note that the cluster combining uses weights to fuse the signals from the RUs $\ell \in \Cc_k(t)$, which replace power control in the UL \cite[Sec. 2.6]{9336188}. 
	It is further shown in \cite[Sec. 7.3]{9336188} that uniform UL power allocation yields comparable results compared to common power control schemes in cell-free networks. 
	We focus on UL results, since by duality, the UL and downlink data rates and system performance are almost identical \cite{9064545, kddi_uldl_precoding, miretti2023ul}. 
	\looseness=-1
	
	\section{Uplink Data Transmission}
	
	Let all active UEs transmit with the same average energy per symbol $P^{\rm ue}$, and we define the system parameter $\SNR \eqdef P^{\rm ue}/N_0$, where $N_0$ denotes the complex baseband noise power spectral density. The received $LM \times 1$ symbol vector at the $LM$ RU antennas for a single channel use on RB $f$ in slot $t$ of the UL is given by
	\begin{equation} 
		\yy (t, f) = \sqrt{\SNR} \; \HH (t, f) \sss (t, f)   + \zz (t, f), \label{ULchannel}
	\end{equation}
	where $\sss (t, f) \in \CC^{K \times 1}$ is the vector
	of information symbols transmitted by the UEs on RB $f$ in slot $t$ (zero-mean unit variance and mutually independent random variables) and 
	$\zz (t, f)$ is an i.i.d. noise vector with components $\sim \Cc\Nc(0,1)$.  
	The goal of cluster $\Cc_k(t)$ is to produce an effective channel observation for symbol $s_k (t, f)$, the $k$-th component of the vector $\sss (t, f)$, from the collectively received signal at the RUs $\ell \in \Cc_k(t)$.  
	Using the receiver vector $\vvv_k (t, f)$,
	the corresponding scalar combined observation for symbol $s_k (t, f)$ is given by 
	$ \hat{s}_k (t, f)  = \vvv_k (t, f)^\herm \yy (t, f). $
	We let $\HH(t) \defines \HH(t, 1:F)$ and $\vvv_k(t) \defines \vvv_k(t, 1:F)$ denote the realization of the channel matrix and of the receiver vector for UE $k$ in time slot $t$ for RBs $f=\{1, \dots, F\}$, respectively.
	The instantaneous mutual information 
	$I( \{ \hat{s}_k(t,f) : f\in [1:F]\} ; \{ s_k(t,f) : f\in [1:F]\})$ in slot $t$
	is a function of $\{ \vvv_k (t), \HH (t)\}$ and given by\footnote{This is in the assumption that 
		the channel state information is known at the receiver and ``Gaussian'' single-user codebooks are used.}
	\begin{gather}
		\Ic_k\left( \vvv_k (t), \HH (t) \right) \defines \frac{1}{F} \sum_{f = 1}^{F} \log\left( 1 + \SINR_k (t, f) \right), \label{eq:ul_mutual_inf}
	\end{gather}
	where we define
	\begin{equation}
		\SINR_k (t, f) = \frac{  |\vvv_k (t, f)^\herm \hh_k (t, f) | ^2 }{ \SNR^{-1}  + \sum_{j \neq k} |\vvv_k (t, f)^\herm \hh_j (t, f) |^2 }.  \label{UL-SINR-unitnorm}
	\end{equation}

	\subsection{Rate Allocation}

	Following \cite{gottsch2022fairness}, we consider outage rates as the effective data rates, such that the receiver can reliably decode an allocated rate under the condition that no information outage occurs \cite{biglieri1998fading}. 
	This condition holds if the allocated rate $r_k$ is smaller than the mutual information $ \Ic_k\left( \vvv_k (t), \HH (t) \right)$.
	The effective instantaneous service rate of UE $k$ in time slot $t$ (expressed in bit per time-frequency channel use) 
	is thus given by \cite{shirani2010mimo}
	\begin{eqnarray}
		\mu_k(t) = \begin{cases}
			(1 - \frac{\tau_p}{T}) R_k(t) ,& \text{if} \; k \in \Ac(t), \\ 
			0 ,&  \text{if } k \notin \Ac(t), \end{cases} 	 \label{eq:allocated_rate}
	\end{eqnarray}
	where, letting $\mathbbm{1} \left\{ \Sc \right\}$ be the indicator function of an event $\Sc$, 
	\begin{eqnarray}
		R_k(t) \defines r_k \times \mathbbm{1} \left\{ r_k < \Ic_k\left( \vvv_k (t), \HH (t) \right) \right\}. \label{eq:def_Rk}
	\end{eqnarray}
	
	Notice that in the information outage regime, even if a user is active (i.e., $k \in \Ac(t)$), it may still have zero rate, depending on 
	the condition $\left\{ r_k < \Ic_k\left( \vvv_k (t), \HH (t) \right) \right\}$. In fact, while the channel state information may be assumed known at the receiver, it is definitely not known at the transmitter, such that instantaneous slot-by-slot 
	rate allocation is not possible. Instead, $r_k$ must be chosen on the basis of
	the random variable $\Ic_k\left( \vvv_k (t), \HH (t) \right)$. In stationary conditions, the instantaneous mutual information distribution is independent of the slot time $t$. In practice, with moderate user mobility, this changes slowly over time. 
	In addition, it is very difficult to analytically characterize such distribution since in general
	$\SINR_k (t, f)$ in (\ref{UL-SINR-unitnorm}) depends not only on the channel state, but also on the set of active users. 
	For the time being, we assume such distribution to be known for each user $k$. Later in this section 
	we present an adaptive algorithmic solution for effective rate allocation. 
	
	The per-user throughput rate is defined as
	\begin{eqnarray}
		\bar{\mu}_k = \lim_{t \rightarrow \infty} \frac{1}{t} \sum_{\tau=0}^{t-1} \mu_k(\tau) 
		= \EE \left[ \mu_k \left( \HH \right) \right] , \label{eq:ue_throughput}
	\end{eqnarray}
	where, with some abuse of notation, we denote by $\mu_k(\HH)$ the random variable induced 
	by the scheduling policy (i.e., the selection of $r_k$ and the active set $\Ac(t)$ in (\ref{eq:allocated_rate}) as a function of the 
	channel state $\HH$), and where the convergence of the time average in (\ref{eq:ue_throughput}) is guaranteed by 
	the ergodicity of the channel process (in our case, i.i.d. over the RBs) and the stationarity of the scheduling policy in the class of dynamic policies considered here \cite{georgiadis2006resource}. 
	
	
	Letting the complementary cumulative distribution function (CDF) of the instantaneous mutual information of
	user $k$ be $P_k(z) \defines \PP( \Ic_k\left( \vvv_k, \HH \right) > z)$,
	from (\ref{eq:def_Rk}) we have that $\EE[R_k(t)] = r_k P_k(r_k)$. Hence, 
	the optimization of $r_k$ is immediate and yields \cite{shirani2010mimo}
	\begin{align}
		r^*_k = \underset{ z }{\arg\max}\ \  z \times P_k(z). \label{eq:r_k_star}
	\end{align}
	Since, as said before, the statistics of $\Ic_k\left( \vvv_k (t), \HH (t) \right)$ for a UE $k$ are 
	generally extremely hard to obtain and depend on the scheduling policy itself, here we consider the 
	localized adaptive scheme proposed in \cite{gottsch2022fairness}, where each user 
	$k$ collects a sliding window of $N$ past samples of the instantaneous mutual information and optimizes its transmission rate 
	$r_k$ using the empirical complementary CDF based on these samples.\footnote{As in 
		\cite{gottsch2022fairness}, the allocated rates are initialized by a ``start-up'' phase consisting of $N_{\rm init}$ time slots. In each of the $N_{\rm init}$ time slots $\Kact$ out of the $\Ktot$ UEs are, considering the conflict graph,  randomly selected to 
		be active. In practice, a user joining the system would start with a very conservative rate and progressively ``ramp up'' 
		the value of $r_k$ until the maximum of the product in (\ref{eq:r_k_star}) is achieved. 
		Actual practical algorithms for rate scheduling work on averaged local statistics along these lines, such that non-stationary (slowly varying) statistics can be tracked.} 
	In the following, we let $\bar{R}_k \defines r_k^* P_k(r^*_k)$, i.e.,  the maximum of the objective function in 
	the right-hand side of (\ref{eq:r_k_star}). 
	
	\section{System Optimization} \label{sec:system_optimization}
	
	We consider a network in the UL with a total number of $\Ktot$ UEs, which operates at its optimal load, when serving $\Kact < \Ktot$ UEs. Further, we assume an {\em infinite backlog} situation, where each UEs has an infinite buffer of data to transmit. 
	By scheduling at most $\Kact$ UEs per time slot, the scheduler wishes to maximize the network utility function, defined as a
	suitable concave component-wise non-decreasing function $g(\cdot)$ of the user throughput rate vector 
	$\bar{\muv} = \left( \bar{\mu}_1, \dots, \bar{\mu}_{\Ktot} \right)$. The problem to be solved is 
	\begin{eqnarray}
		& \text{maximize} & g(\bar{\muv}) , \ \ \ \ \text{subject to } \bar{\muv} \in \Rs ,  \label{eq:max_utility} 
	\end{eqnarray}
	where $\Rs$ is the system achievable throughput rate region \cite{georgiadis2006resource}. 
	Since $\Rs$ is not characterized easily, the solution $\bar{\muv}^\star$ of (\ref{eq:max_utility}) is generally very hard to 
	find analytically \cite{shirani2010mimo}. However, the framework of \cite{georgiadis2006resource} can be used to find a scheduling scheme that approximates $\bar{\muv}^\star$ within any desired accuracy. 
	
	Specifically, the scheduler solves at each scheduling slot $t$ the 
	weighted sum rate maximization (WSRM) problem (with respect to the active set $\Ac(t)$)
	\begin{equation} 
		\max \;\; \sum_{k\in \Ac(t)} Q_k(t) \bar{R}_k, \label{wsrm}
	\end{equation}
	where $\{Q_k(t)\}$ are the backlogs of ``virtual queues'' used as weights in (\ref{wsrm}) with update rule
	\begin{align}
		Q_k(t+1) = \max\left\{ Q_k(t) - \mu_k(t), 0 \right\} + A_k(t)  \label{eq:queue_update}
	\end{align}
	and $\{A_k(t)\}$ is a set of ``virtual arrival processes''.
	For each $t$, we have $A_k(t) = a_k$, where $\av = (a_1, \dots, a_{\Ktot})$ is the solution to the convex optimization problem
	\begin{equation}
		\begin{array}{l l}
			\underset{\av}{\text{maximize}} & \ Vg(\av) - \sum_{k \in [\Ktot]} Q_k\left( t \right) a_k  \\
			\text{subject to} & \ 0 \leq a_k \leq A_{\rm max}, \ \ \forall k \in [\Ktot]. 
		\end{array} \label{eq:opt_ak_general}
	\end{equation}
	Here, $V$ and $\ A_{\rm max}$ are suitably chosen constant parameters that determine the behavior of the algorithm \cite{georgiadis2006resource}. In particular, it is known that for $A_{\max}$ sufficiently large the time-averaged service rates generated by the algorithm (i.e., $\frac{1}{t} \sum_{\tau = 1}^t \mu_k(\tau)$ for large $t$) approximate the optimal throughput rate point solution of (\ref{eq:max_utility}) within a gap $O(1/V)$, while the 
	time-averaged sum queue backlog grows as $O(V)$.\footnote{The proof under the assumptions made in this paper differs from the performance guarantees given in \cite{shirani2010mimo,georgiadis2006resource} and will be published in a journal paper on scheduling in cell-free massive MIMO. Further, in most literature, the time-averaged queue backlog is referred to as ``delay'' but here since we are in the infinite buffer regime and the queues are virtual, 
	this quantity is rather an indication of the time it takes for the algorithm to converge to a stationary state. \looseness=-1}
	
	\subsection{Proportional Fairness and Hard Fairness Scheduling}
	
	We consider PFS and HFS, leading to different solutions to the optimization problem (\ref{eq:opt_ak_general}).
	In case of PFS, we have $g(\av) = \sum_{k \in [\Ktot]} \log a_k$ in (\ref{eq:opt_ak_general}),
	which yields the arrivals \cite{shirani2010mimo}
	\begin{equation}
		a_k = \min \left\{ \frac{V}{Q_k(t)}, A_{\rm max} \right\} . \label{eq:a_k_pfs}
	\end{equation}
	For HFS, $g(\av) = \underset{k \in [\Ktot]}{\min} \ a_k$ and the solution to (\ref{eq:opt_ak_general}) is \cite{shirani2010mimo}
	\begin{align}
		a_k = \begin{cases}  A_{\rm max} , & \text{if } V > \sum_{k \in [\Ktot]} Q_k(t), \\ 
			0 , & \text{else. }  \end{cases} \label{eq:a_k_hf}
	\end{align}
	
	\section{Algorithmic Solutions} \label{algorithmic}
	
	We first describe an algorithmic solution proposed in \cite{gottsch2022fairness} including a reassignment of pilots and clusters at each scheduling decision. Then we will describe our proposed scheme with fixed pilot and cluster assignments, reducing greatly the required communication between UEs and RUs. For both schemes, $\Qm(0) = \zerov$, and UEs with empty queues are not scheduled, so in some slots the number of active UEs may be smaller than $\Kact$, in particular when HFS is employed.
	
	\subsection{Pilot and Cluster Reallocation Scheme}
	
	This scheme proposed in \cite{gottsch2022fairness} carries out the UL pilot allocation for channel estimation and cluster formation in each time slot after selecting the set of active UEs. 
	Having defined $\Kact$ as the desired number of simultaneously active UEs, we solve the WSRM (\ref{wsrm}) as
	\begin{equation}
		\begin{array}{l l}
			\underset{\xv}{\rm maximize} & \ \ \sum_{k \in [\Ktot]} Q_k(t) \bar{R}_k x_k  \\
			\text{subject to} & \ \ \sum_{k \in [\Ktot]} x_k \leq \Kact ,  \\
			& \ \ x_k \in \left\{ 0, 1 \right\} , \label{eq:sched_asilomar}
		\end{array}
	\end{equation}
	where $x_k=1$ if UE $k \in \Ac(t)$ and $0$ otherwise. The solution is immediate and consists of sorting the users in decreasing order of the product 
	$Q_k(t) \bar{R}_k$ and letting $\Ac(t)$ the set of the top $\Kact$ sorted users. 
	Given the selected set $\Ac(t)$, UL pilots and user-centric clusters are assigned to the active UEs 
	following the semi-overloaded pilot assignment method from \cite{osawa2022effective}, where an RU may assign the same pilot sequence to multiple UEs provided that the channel subspaces of the UEs are nearly mutually orthogonal, 
	such that accurate channel estimation is possible (negligible mutual pilot contamination, using the decontamination method of  \cite{gottsch2022subspace}).  An RU-UE association can only be established, when the SNR association threshold criterion $\beta_{\ell,k} \geq \frac{\eta}{M \SNR } $ is fulfilled, where $\eta$ is an association threshold parameter. 
	
	\subsection{Fixed Pilots and Clusters}
	
	In this case, we first define a conflict graph $\Cs = ([\Ktot], \Ec_\Cs) $ with a vertex set corresponding to all $\Ktot$ UEs in the network and an edge set $\Ec_\Cs$ accounting for the conflicts.
	Letting $p_k$ denote the UL pilot index of UE $k$, we define that a UE-pair $(k, k')$ has a scheduling conflict if 
	\begin{enumerate}[leftmargin=*]
		\item the UEs are associated to at least one common RU, i.e., $\Cc_{k, k'} \defines \Cc_k \cap \Cc_{k'} \neq \emptyset$, and 
		\item the UEs are assigned the same UL pilot, i.e., $p_k = p_{k'}$, and 
		\item the subspaces of the UEs overlap with regard to at least one RU $\ell \in \Cc_{k, k'}$, i.e., $\sum_{\ell \in \Cc_{k, k'}} \left| \Sc_{\ell,k} \cap \Sc_{\ell,k'} \right| > 0$, 
	 where $| \cdot | $ denotes the cardinality of a set.
	\end{enumerate}
	The graph $\Cs$ has an edge between the vertex $k$ and vertex $k'$ for all UE-pairs $(k, k')$ in conflict,
	with the meaning that any UE-pair $(k, k') \in \Ec_\Cs$ is not allowed to be scheduled in the same time slot, since they would interfere in the channel estimation process.
	Based on this conflict definition, we propose the following pilot assignment and cluster formation scheme. 
	\begin{enumerate}[leftmargin=*]
		\item When a UE $k$ joins the system, it connects to a maximum of $Q$ RUs with the largest LSFCs, provided that $\beta_{\ell,k} \geq \frac{\eta}{M \SNR } $, forming the set $\Cc_k$.\footnote{In a practical system, UEs join and leave the network, such that each UE could be assigned a pilot and an RU cluster according to the proposed scheme. In our simulations, we carry out the proposed scheme for each UE in the order of their index.}
		\item For each UL pilot index $\tau^{(i)} = [\tau_p ]$ the RUs $\ell \in \Cc_k$ count the number of associated UEs $k' \neq k: k' \in \Uc_\ell$ with a non-orthogonal subspace. The set of potentially conflicting UEs with pilot $\tau^{(i)}$ regarding UE $k$ is given by $ \Cs_k (\tau^{(i)} ) = \left\{ \bigcup_{\ell \in \Cc_k} \Cs_{\ell, k} (\tau^{(i)} ) \right\}$, where 
		\begin{eqnarray}
			\Resize{1\linewidth}
			{
				\Cs_{\ell, k} (\tau^{(i)} ) = \left\{ k' \in \Uc_\ell : \right. \left.
				\mathbbm{1} \{ p_{k'} = \tau^{(i)} \} \cap \mathbbm{1} \left\{ \left| \Sc_{\ell,k} \cap \Sc_{\ell,k'} \right| > 0 \right\} \right\} , \nonumber 
			}
		\end{eqnarray}
		and where $\Cs_k (\tau^{(i)} )$ contains each possible UE only once. Here, perfect subspace knowledge at RU $\ell$ for associated UEs $k \in \Uc_\ell$ is assumed. Schemes for channel subspace and covariance matrix estimation, respectively, in cell-free massive MIMO are presented in \cite{gottsch2022subspace, 9715152}.
		\item The pilot corresponding to the smallest number of conflicting UEs, i.e., $\tau^{(i^\star)} = \underset{ i }{\arg \min} \ | \Cs_k (\tau^{(i)}) |  $, is assigned to UE $k$. If more than one pilot corresponds to the smallest number of conflicting UEs, an arbitrary choice of these pilots is made.
	\end{enumerate}
	The fixed pilot and cluster assignment to all $\Ktot$ UEs in the network is carried out independently of scheduling decisions. 
	The resulting WSRM problem (\ref{wsrm}), subject to the conflict graph, is given by the linear integer program 
	\begin{equation}
		\begin{array}{l l}
			\underset{\xv}{\text{maximize}} & \ \ \sum_{k \in [\Ktot]} Q_k(t) \bar{R}_k x_k  \\
			\text{subject to} & \ \ \sum_{k \in [\Ktot]} x_k \leq \Kact ,  \\
			& \ \ x_k \in \left\{ 0, 1 \right\} , \\
			& \ \ x_k + x_{k'} \leq 1 , \ \forall (k, k') \in \Ec_\Cs, \label{eq:sched_conflict_graph}
		\end{array}
	\end{equation}
	which can be efficiently solved with standard tools (e.g., Gurobi or MATLAB) even for fairly large systems.
	 
	\section{Numerical Evaluations and Outlook}
	\begin{figure}[t!]
		\centering
		\begin{tikzpicture}[define rgb/.code={\definecolor{mycolor}{RGB}{#1}},
			rgb color/.style={define rgb={#1},mycolor}]
			\begin{axis}[
				enlarge y limits=false,clip=false,
				enlarge x limits=false, clip=true,
				no markers,
				width=.52\linewidth, 
				height=3.8cm, 
				grid=major, 
				grid style={gray!30}, 
				xlabel= {\small $\bar{\mu}_k$}, 
				ylabel={\small Empirical CDF},
				x unit=\si{bit/s/Hz}, 
				yticklabels={},
				extra y ticks={0, .1, .2, .3, .4, 0.5, .6, .7, .8, .9, 1},
				extra y tick labels={0,,,,,0.5,,,,,1},
				xticklabels={},
				extra x ticks={0, .1, .2, .3, .4, 0.5, .6},
				extra x tick labels={0,,0.2,,0.4,,0.6},
				x tick label style={yshift=-2,rotate=0,anchor=north}, 
				xlabel style={yshift=4,rotate=0,anchor=north}, 
				ticklabel style={font=\small},
				xmin=0,
				xmax=.6,
				ymin=0,
				ymax=1,
				legend style={row sep=-0.07cm, inner ysep = 0cm, inner xsep = 0.05cm, at={(0.01,0.02), font=\fontsize{7}{12}}, anchor=south west},
				legend cell align={left},
				legend style={sehr kurze Legende}]
				\addplot[rgb color={0, 114, 189}, line width=0.6pt]
				table[blue,col sep=comma] {csvs/hf_adaptive_V10000.csv}; 			
				\addlegendentry{{ Baseline}};
				\addplot[rgb color={217, 83, 25}, line width=0.6pt]
				table[blue,col sep=comma] {csvs/hf_fixed_re_V10000.csv}; 			
				\addlegendentry{{Proposed}};
			\end{axis}
		\end{tikzpicture}
		\begin{tikzpicture}[define rgb/.code={\definecolor{mycolor}{RGB}{#1}},
			rgb color/.style={define rgb={#1},mycolor}]
			\begin{axis}[
				enlarge y limits=false,clip=false,
				enlarge x limits=false, clip=true,
				no markers,
				width=.52\linewidth, 
				height=3.8cm, 
				grid=major, 
				grid style={gray!30}, 
				xlabel= {\small $\bar{\mu}_k$}, 
				ylabel={\small Empirical CDF},
				x unit=\si{bit/s/Hz}, 
				yticklabels={},
				extra y ticks={0, .1, .2, .3, .4, 0.5, .6, .7, .8, .9, 1},
				extra y tick labels={0,,,,,0.5,,,,,1},
				xticklabels={},
				extra x ticks={0, .1, .2, .3, .4, 0.5, .6, .7, .8, .9, 1, 1.1},
				extra x tick labels={0,,,,,0.5,,,,,1,},
				x tick label style={yshift=-2,rotate=0,anchor=north}, 
				xlabel style={yshift=4,rotate=0,anchor=north}, 
				ticklabel style={font=\small},
				xmin=0,
				xmax=1.2,
				ymin=0,
				ymax=1,
				legend style={row sep=-0.07cm, inner ysep = 0cm, inner xsep = 0.05cm, at={(0.99,0.02), font=\fontsize{7}{12}}, anchor=south east},
				legend cell align={left},
				legend style={sehr kurze Legende}]
				\addplot[rgb color={0, 114, 189}, line width=0.6pt]
				table[blue,col sep=comma] {csvs/ppf_adaptive_V10000.csv}; 			
				\addlegendentry{{ Baseline}};
				\addplot[rgb color={217, 83, 25}, line width=0.6pt]
				table[blue,col sep=comma] {csvs/ppf_fixed_V10000.csv}; 			
				\addlegendentry{{Proposed}};
			\end{axis}
		\end{tikzpicture}
		\vspace{-.8cm}
		\caption{User throughput for HFS (left) and PFS (right), where ``Baseline'' accounts for the reassignment scheme from \cite{gottsch2022fairness}.}
		\label{fig:user_tp_cdf_hfs_pfs_clustering}
		\vspace{-.6cm}
	\end{figure}
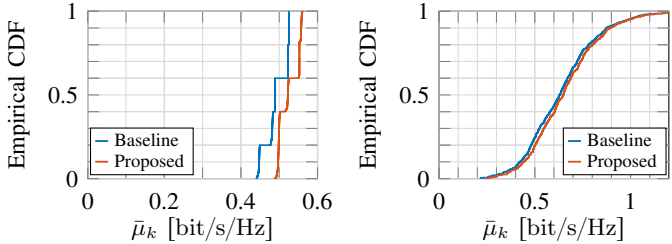
	We consider a system like in \cite{gottsch2022fairness}, i.e., a squared network area of $A = 50 \times 50 \text{m}^2$ with a torus topology to avoid boundary effects, containing $L=12$ RUs, each with $M=8$ antennas, and $K=100$ UEs. 
	We assume a bandwidth of $W= 10$ MHz and noise with power spectral density of $N_0 = -174$ dBm/Hz.
	We let the angular interval of length $\Delta = \pi/8$, the SNR threshold $\eta = 1$ and the maximum cluster size $Q=10$ (RUs serving one UE) in the simulations.
	We define $P^{\rm ue}$ such that $\bar{\beta} M \SNR = 1$ (i.e., 0 dB), when the expected pathloss $\bar{\beta}$ with respect to LOS and NLOS is calculated for distance $3 d_L$, where $d_L = \sqrt{\frac{A}{\pi L}}$ is the radius of a disk of area equal to $A/L$. This leads to a certain level of overlap of the RUs' coverage areas considering the SNR association threshold. The UEs are randomly dropped in the network area, while the RUs are placed on $3\times 4$ rectangular grid. The online rate adaptation is carried out for all schemes with  $N_{\rm init} = 500$ and $N = 100$, and we consider RBs of dimension $T = 200$ symbols. 
	Since we consider a network like in \cite{gottsch2022fairness}, we choose $\tau_p = 20$ and $\Kact=40$, the approximate optimal parameters according to the results in \cite{gottsch2022fairness}. 
	We simulate 5 different setups (random placement of UEs) that are equal for all investigated approaches. The algorithm parameters are chosen as $A_{\rm max} = 100$ and $V=10000$. We refer the reader to \cite{shirani2010mimo, gottsch2022fairness} for an evaluation of different values of $V$.
	\subsection{Fixed Pilots and Clusters vs. Reassignment Scheme}
	Considering a narrowband system with $F=1$ RB,
	Fig. \ref{fig:user_tp_cdf_hfs_pfs_clustering} shows that the proposed method with fixed pilot and cluster allocations (slightly) outperforms the reassignment scheme. The proposed method has the advantage that each UE $k$ is connected to all (at most $Q$) RUs with the largest LSFCs. Severe pilot contamination is then prevented by the conflict graph.
	In contrast, a scheduled UE $k$ using the reassignment scheme might end up being connected to only a fraction of possible serving RUs. This can happen since conflicts are avoided by not associating a UE $k$ to a possible RU $\ell$ if that RU already serves another UE $k'$ with the same pilot and a non-orthogonal channel subspace, i.e., $\left| \Sc_{\ell,k} \cap \Sc_{\ell,k'} \right| > 0$. In this way however, severe interference is not prevented.
	\subsection{PFS in a Wideband System}
	\begin{figure}[t!]
		\centering
		\begin{tikzpicture}[define rgb/.code={\definecolor{mycolor}{RGB}{#1}},
			rgb color/.style={define rgb={#1},mycolor}]
			\begin{axis}[
				enlarge y limits=false,clip=false,
				enlarge x limits=false,
				no markers,
				width=.54\linewidth, 
				height=3.4cm, 
				grid=major, 
				grid style={gray!30}, 
				ylabel={\footnotesize Empirical CDF},
				label style={inner sep=0pt}, 
				legend style={at={(0.5,-0.2)},anchor=north}, 
				yticklabels={},
				extra y ticks={0, .1, .2, .3, .4, 0.5, .6, .7, .8, .9, 1},
				extra y tick labels={0,,,,,0.5,,,,,1},
				xticklabels={},
				extra x ticks={0, 1, 2, 3, 4, 5},
				extra x tick labels={0, 1, 2, 3, 4, 5},
				x tick label style={yshift=-2,rotate=0,anchor=north}, 
				tick label style={font=\footnotesize },
				xmin=0,
				xmax=5,
				ymin=0,
				ymax=1,
				legend pos=south east,
				legend style={row sep=-0.12cm, inner ysep = 0cm, inner xsep = 0.05cm, font=\fontsize{7}{12}, at={(0.995,0.02)}, anchor=south east},
				legend cell align={left},
				legend style={sehr kurze Legende}]
				\addplot[rgb color={0, 114, 189}, line width=0.6pt]
				table[blue,col sep=comma] {csvs/data_mut_info_wideband.csv}; 			
				\addlegendentry{{ $F=8$}};
				\addplot[rgb color={217, 83, 25}, line width=0.6pt]
				table[blue,col sep=comma] {csvs/data_mut_info_narrowband.csv}; 			
				\addlegendentry{{ $F=1$}};
				\begin{scope}[scale=0.8, transform shape]
					\draw (5.8, -0.46) node[left,black]{$\Ic_k\left( \vvv_k { , } \HH \right)$ \si{[bit/s/Hz]}};
				\end{scope}
			\end{axis}
		\end{tikzpicture}
		\begin{tikzpicture}[define rgb/.code={\definecolor{mycolor}{RGB}{#1}},
			rgb color/.style={define rgb={#1},mycolor}]
			\begin{axis}[
				enlarge y limits=false,clip=false,
				enlarge x limits=false,
				no markers,
				width=.54\linewidth, 
				height=3.4cm, 
				grid=major, 
				grid style={gray!30}, 
				ylabel={\footnotesize Empirical CDF},
				label style={inner sep=0pt}, 
				legend style={at={(0.5,-0.2)},anchor=north}, 
				yticklabels={},
				extra y ticks={0, .1, .2, .3, .4, 0.5, .6, .7, .8, .9, 1},
				extra y tick labels={0,,,,,0.5,,,,,1},
				xticklabels={},
				extra x ticks={0, 0.5, 1, 1.5, 2},
				extra x tick labels={0, 0.5, 1, 1.5, 2},
				x tick label style={yshift=-2,rotate=0,anchor=north}, 
				ticklabel style={font=\footnotesize },
				xmin=0,
				xmax=1.5,
				ymin=0,
				ymax=1,
				legend pos=south east,
				legend style={row sep=-0.12cm, inner ysep = 0cm, inner xsep = 0.05cm, font=\fontsize{7}{12}, at={(0.995,0.02)}, anchor=south east},
				legend cell align={left},
				legend style={sehr kurze Legende}]
				\addplot[rgb color={0, 114, 189}, line width=0.6pt]
				table[blue,col sep=comma] {csvs/data_tp_wideband_cut_to_1p5_on_xaxis.csv}; 			
				\addlegendentry{{ $F=8$}};
				\addplot[rgb color={217, 83, 25}, line width=0.6pt]
				table[blue,col sep=comma] {csvs/data_tp_narrowband.csv}; 			
				\addlegendentry{{ $F=1$}};
				\begin{scope}[scale=0.77, transform shape]
					\draw (1.83, -.46) node[left,black, text opacity=0]{$\Ic_k\left( \vvv_k { , } \HH \right)$ \si{[bit/s/Hz]}};
				\end{scope}
				\begin{scope}[scale=0.8, transform shape]
					\draw (1.48, -.46) node[left,black, text opacity=1]{$\bar{\mu}_k$ \si{[bit/s/Hz]}};
				\end{scope}
			\end{axis}
		\end{tikzpicture}
		\vspace{-.3cm}
		\caption{The empirical CDF of $\Ic_k( \vvv_k, \HH )$ of an example UE $k$ (left) and the user throughput (right) with $F = \{1, 8\}$ in a network employing PFS.}
		\label{fig:empirical_cdf_narrow_wideband}
		\vspace{-.5cm}
	\end{figure}
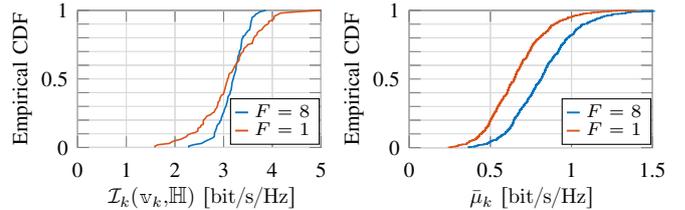
	We compare the performance of a wideband system with $F=8$ RBs 
	to the narrowband system with $F=1$ using the proposed pilot and cluster allocation method.
	Fig. \ref{fig:empirical_cdf_narrow_wideband} shows that because of coding over $F=8$ RBs in (\ref{eq:ul_mutual_inf}), the empirical CDF of the instantaneous mutual information behaves in a more deterministic way. This allows a more aggressive instantaneous rate allocation in the active slots. As a result, see Fig. \ref{fig:empirical_cdf_narrow_wideband}, the user throughput rate in a system with $F=8$ can be significantly increased compared to $F=1$. 
	
	\subsection{Concluding Remarks}
	In this work, we considered a user-centric cell-free massive MIMO network with a total number of users in its area that is much larger than the optimal user load. 
	We proposed a fixed pilot and cluster assignment scheme under scheduling, which greatly reduces the communication overhead between UEs and RUs, while also achieving better performance compared to the scheme in \cite{gottsch2022fairness}. 
	We further showed that when coding over several RBs in a wideband system, the mutual information can be predicted more accurately, yielding a smaller probability of information outage. This in turn leads to a larger UE throughput compared to the narrowband system with one RB. 
	
	
	
	\bibliography{IEEEabrv,isit23}
	
	%
	%

\end{document}